\documentclass[letterpaper,english,aps,prb,reprint,superscriptaddress]{revtex4-2}
\usepackage[T1]{fontenc}
\usepackage[latin9]{inputenc}
\usepackage{xcolor}
\usepackage{mathrsfs}
\usepackage{amsmath}
\usepackage{amssymb}
\usepackage{makeidx}
\makeindex
\usepackage{graphicx}

\makeatletter

\usepackage{hyperref}
\hypersetup{
    colorlinks = true,
	allcolors = blue	
}
\usepackage{times}

\makeatother

\usepackage{babel}
\begin{document}
\bibliographystyle{apsrev4-2.bst}
\title{Magnetization-Induced Phase Transitions on the surface of 3D
Topological Insulators }
\author{Yu-Hao Wan}
\affiliation{International Center for Quantum Materials, School of Physics,
Peking University, Beijing 100871, China}
\author{Qing-Feng Sun}
\thanks{Corresponding author: sunqf@pku.edu.cn.}
\affiliation{International Center for Quantum Materials, School of Physics,
Peking University, Beijing 100871, China}
\affiliation{Hefei National Laboratory, Hefei 230088, China}

\begin{abstract}
From the low-energy model, the topological field theory
indicates that the surface magnetization can open a surface
gap in 3D topological insulators (TIs),
resulting in a half-quantized Hall conductance.
Here by employing the realistic lattice model, we show the occurrence of
{ the surface phase transitions}, accompanied with the
sharp changes of the surface Chern number from $\frac{1}{2}$
to $-\frac{3}{2}$ finally to $-\frac{1}{2}$,
in 3D TIs induced by surface magnetization.
These surface phase transitions lead to the sudden jumps in the
magneto-electric coefficient and the quantum Hall conductance,
which are experimentally observable.
Furthermore, we present the phase diagram that elucidates
the behavior of the 3D TI surface Chern numbers under
surface magnetization for different $Z_{4}$ topological numbers.
Our study highlights the presence of the new phases
with broken bulk-boundary correspondence and
enriches understandings of the properties of TIs.
\end{abstract}
\maketitle

\section{\label{sec:level1}Introduction}
Three-dimensional (3D)
topological insulators (TIs), featured by a bulk band gap
and gapless helical surface states protected
by the time-reversal symmetry (TRS), { have} received intense
research over the past decade \citep{bansil2016emphcolloquium,hasan2010emphcolloquium,bernevig2006quantum,kane2005quantum,konig2007quantum}.
The physics origin of the gapless surface states is
well known as bulk-boundary correspondence,
in which the gapless surface states are determined
by the topology of the bulk band.
When the TRS is broken on the surface (e.g. by the magnetization),
the fall of the bulk-boundary correspondence
will result in a gapped surface.
Nevertheless, even if on the TRS-breaking surface,
information about the bulk's topological properties
can still be manifested at the gapped surface.
For instance, half-quantized anomalous Hall { conductance}(AHC)
and topological magneto-electric (ME) effects \citep{wan2022topological,wang2015quantized}
can arise within the gap of the TRS-breaking surface \citep{essin2009magnetoelectrica,mogi2022experimental,sekine2021axionelectrodynamics,nenno2020axionphysics,qi2008topological,fu2007topological,zhou2022transport},
which relates with the topological $\theta$-axion term in the bulk
\citep{wilczek1987twoapplications}.
In 3D TIs, the $\theta$-term is directly associated with the $Z_{2}$
topological index, where $\theta=\pi$ (mod 2\ensuremath{\pi})
corresponds to a nontrivial bulk \citep{qi2008topological}.
It is worth noting that the bulk-boundary correspondence
on the TRS-breaking surface, and then the half-quantized
AHC and the topological ME effect, are considered outcomes
of low-energy models near $\Gamma$ point
based on topological field theory \citep{qi2011topological,qi2008topological}.
For a strong magnetization with the TRS strongly broken,
the availability of low-energy models is poorly studied.
In reality, the system should be described on a complete Brillouin zone.
The low-energy model fails to capture information from the Brillouin zone,
thereby missing certain physics phenomena related to it.

\begin{figure}
    \begin{centering}
    \includegraphics[scale=0.44]{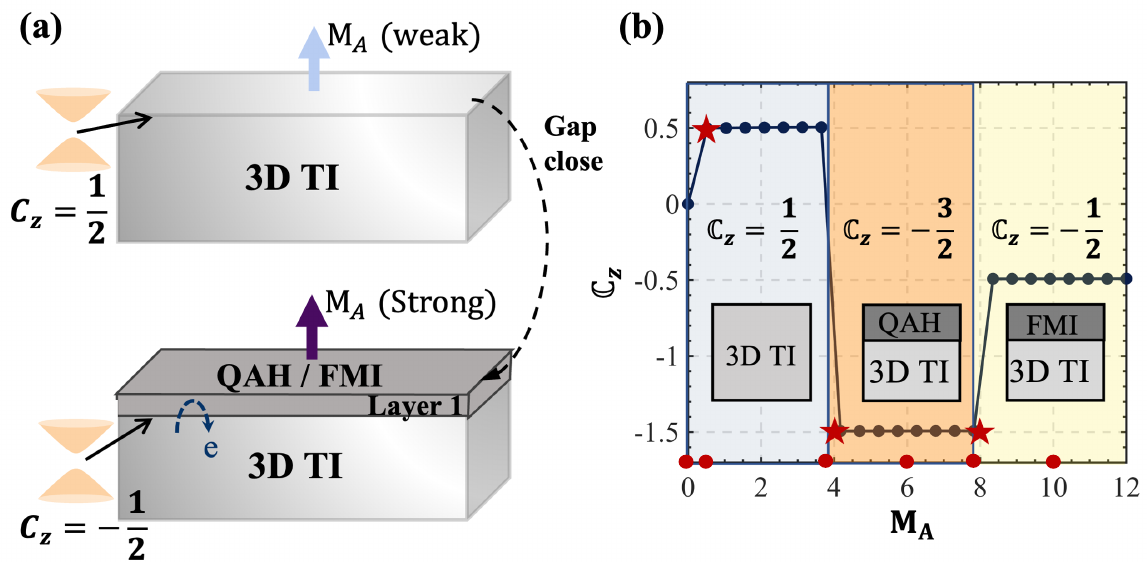}
    \par\end{centering}
    \caption{\label{fig:1} (a) Upper and lower parts are schematic {diagrams}
    for the surface phase of 3D TI under the weaker and stronger
    surface magnetization, respectively.
    (b) Surface Chern number $\mathbb{C}_{z}$ as a
    function of surface magnetization $M_A$.
    The blue, orange, and yellow regions {correspond} to
    $\mathbb{C}_{z}= \frac{1}{2}$, $-\frac{3}{2}$, and $-\frac{1}{2}$.
    The insets illustrate schematic diagrams of the systems in the three regions.}
    \end{figure}

In this paper, we systematically investigate the influence of
surface magnetization on 3D TIs within the realistic lattice model.
We discover that the increasing magnetization leads to a series of
topological phase transitions on the surface of the 3D TI.
Specifically, the surface Chern number transitions
from $\frac{1}{2}$ to $-\frac{3}{2}$ and eventually
reaches $-\frac{1}{2}$ [see Fig. \ref{fig:1}(b)].
To quantitatively characterize these transitions,
we employ a slab model and an effective three-layer
model, calculating the local Chern markers from
both numerical and analytical perspectives.
The computational results reveal that surface magnetization initiates
a non-adiabatic transformation in the first layer of the 3D TI,
converting the massive Dirac fermion into a state of
quantum anomalous Hall/ferromagnetic insulator (QAH/FMI).
Furthermore, as the first-layer topological transition occurs,
an opposing massive Dirac fermion reemerges in the second layer
and results in $-\frac{1}{2}$ Chern number [see Fig. \ref{fig:1}(a)].
Moreover, these topological phase transitions also lead to
the {sharp} changes of both topological ME coefficient
and the surface Hall conductance plateau.

The organization of this paper is as follows. In Sec. \ref{sec:level2}, we present our lattice model and numerically investigate the variation of the surface Chern number with the increase of surface magnetization using a 20-layer 3D TI slab model. In Sec. \ref{sec:level3}, we establish an effective 3-layer model to describe the behavior of 3D TI surface states in the presence of surface magnetization. In Sec. \ref{D}, we present a phase diagram that illustrates how different bulk topological mass terms affect the surface topological transitions induced by surface magnetization.
Both the topological ME coefficient and
the surface quantum Hall conductance plateaus
induced by the surface topological transitions are studied in
Sec. \ref{sec:level4}.
We discuss experimental implementation in Sec. \ref{sec:level5} and conclude with a summary. Additional computational details and supplementary figures can be found in Appendices \ref{C} to \ref{real_mat}.

\section{\label{sec:level2}Surface magnetization-induced phase transitions}

\begin{figure}
    \begin{centering}
    \includegraphics[scale=0.5]{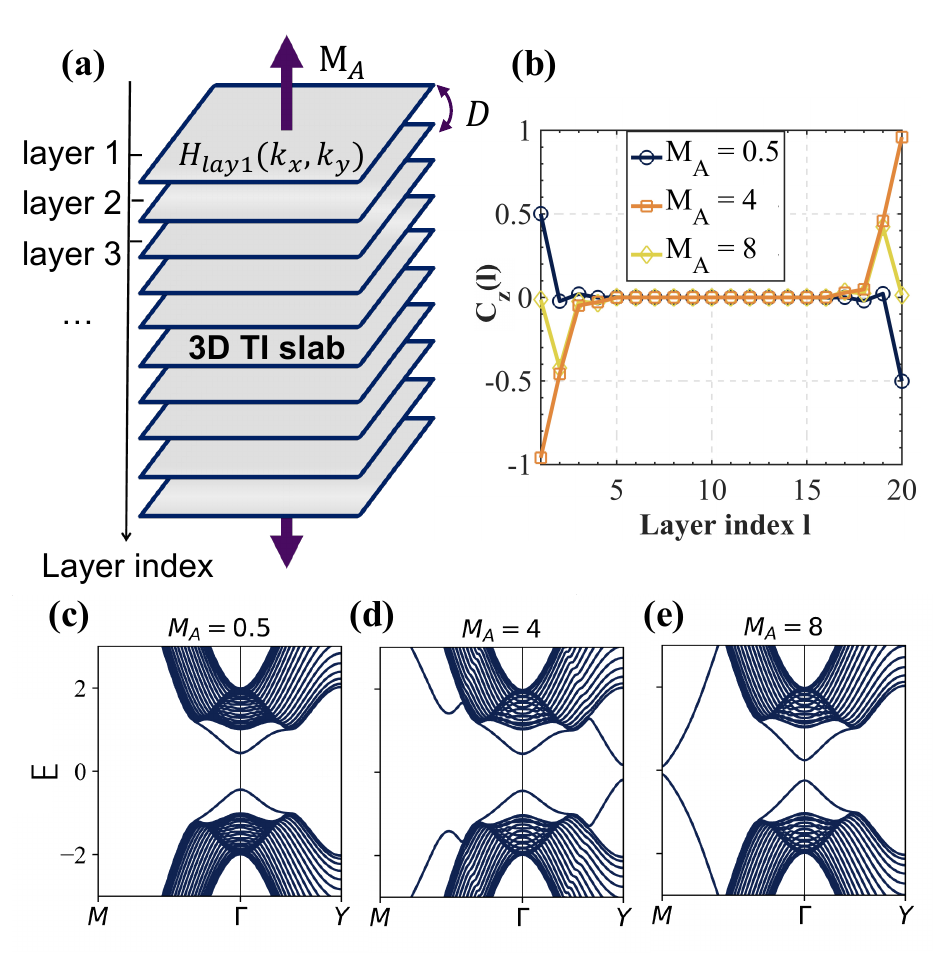}
    \par\end{centering}
    \caption{\label{fig:2} (a) Schematic diagram of a 20-layer slab model
    of a 3D TI, with opposite magnetization $\ensuremath{{M_{A}}}$
    applied to the upper and lower surfaces. $D$ denotes inter-layer hopping.
    (b) Layer-resolved local Chern markers $C_z(l)$ of a 20-layer 3D TI slab model with different $M_A$ indicated by pentagrams
    in Fig. \ref{fig:1}(b). (c) Band structures under different $M_A$.}
    \end{figure}

We start with the Hamiltonian of a 3D TI in cubic lattice:
\begin{equation}
\mathcal{H}_{\mathrm{TI}}=
(m-6B)s_{0}\sigma_{z} +
\sum_{i}(2B s_{0}\sigma_{z}\cos k_{i}+A s_{i}\sigma_{x} \sin k_{i})
\label{eq:1}
\end{equation}
where $m$, $B$ and $A$ are the model's parameters and
$k_{i}$ is the momentum with $i=x,y,z$.
$s_{i}$ and $\sigma_{i}$ are the Pauli matrices on the spin and orbital spaces.
The system exhibits a nontrivial topological phase when satisfying
$0<m<4B$ or $8B<m<12B$ \citep{shen2012topological}.
Below, we choose $m=2$, $B=1$ and $A=1$ to position the
system in the nontrivial topological phase, unless mentioned otherwise.
{ We construct a 20-layer thick slab model  [Fig. \ref{fig:2}(a)]. For the convenient observation of the influence of surface magnetization on energy bands, we introduce symmetrical magnetization on both the top and bottom surfaces, thereby opening magnetic energy gaps on both surfaces. }
The Hamiltonian is represented as follows:
\[
\ensuremath{{H_{slab}}=\left({\begin{array}{ccccc}
{{H_{lay_{1}}}} & D & 0 & ... & 0\\
{D^{\dagger}} & {H_{lay_{2}}} & D &...& 0\\
0 & {D^{\dagger}} & {H_{lay_{3}}} & {...} & 0\\
... & ... & ... & {...} & ...\\
0 & 0 & 0 & ... & {{H_{lay_{20}}}}
\end{array}}\right)}
\]
The Hamiltonian for each layer is given by
$\ensuremath{{H_{lay_l}}}=\ensuremath{{H_{lay}}} + \delta_{l,1} M_A s_z
-\delta_{l,20} M_A s_z$
with $\ensuremath{{H_{lay}}=A {\sigma_{x}}{s_{x}} \sin{k_{x}}+
A{\sigma_{x}}{s_{y}} \sin{k_{y}}+
[{m-6B+2B(\cos{k_{x}}+\cos{k_{y}}})]{\sigma_{z}}{s_{0}}}$.
The interlayer hopping is given by $\text{\ensuremath{D=\frac{A}{{2i}}{\sigma_{x}}{s_{z}}+B{\sigma_{z}}{s_{0}}}}$.
$M_{A}$ represents the magnetization strength of the top and bottom
surfaces, which can be introduced experimentally through magnetic
doping or heterostructure approaches \citep{mogi2022experimental,xiao2018realization,mogi2017amagnetic,yasuda2017quantized}.

Then we investigate the local Chern markers within
the slab model to characterize the contributions of different layers
to the total Chern number \citep{bianco2011mapping,rauch2018geometric,varnava2018surfaces}.
The local Chern marker projected onto layer $l$, denoted as $\ensuremath{{C_{z}}\left(l\right)}$, can be calculated from the
expression \cite{varnava2018surfaces}:
\begin{equation}
\ensuremath{{C_{z}}(l)=\frac{{-4\pi}}{\mathscr{A}}{\mathop{{\rm Im}}\nolimits}\frac{1}{{N_{k}}}\sum\limits _{k}{\sum\limits _{v{v^{\prime}}c}{X_{vck}}}Y_{{v^{\prime}}ck}^{\dagger}{\rho_{v{v^{\prime}}k}}(l)}
\label{eq:2}
\end{equation}
The matrix element for the position operator along the \ensuremath{x} or \ensuremath{y}
directions, is denoted as $\ensuremath{X{\left(Y\right)_{vc{\bf {k}}}}=\left\langle {\psi_{v{\bf {k}}}}\right|x\left(y\right)\left|{\psi_{c{\bf {k}}}}\right\rangle =\frac{{\left\langle {\psi_{v{\bf {k}}}}\right|i\hbar{v_{x}}\left({v_{y}}\right)\left|{\psi_{c{\bf {k}}}}\right\rangle }}{{{E_{c{\bf {k}}}}-{E_{v{\bf {k}}}}}}},$
which is related to the energy difference between the conduction
and valence bands ${E_{c{\bf {k}}}}-{E_{v{\bf {k}}}}$.
The indices \ensuremath{v} and \ensuremath{c} represent the valence and conduction
bands. $\ensuremath{\rho_{vv'k}\left(l\right)}$
is the projection matrix on to the corresponding layer \ensuremath{l}
which implies a summation over all orbitals $v, v', c$ belonging to that layer. $N_{k}$ represents
the number of \ensuremath{k}-points and $\mathscr{A}$ represents the unit cell area.

In Fig. \ref{fig:2}(b) and (c-e),
we have computed the local Chern markers and corresponding
band structures under some representative magnetizations,
which are marked with red pentagrams in Fig. \ref{fig:1}(b).
{Due to the opposite magnetization  on the top and bottom surfaces, the local Chern marker on these surfaces is also opposite (see Fig. \ref{fig:2}(b)). As a result, the overall Chern number of the system is zero, leading to the absence of a net Hall effect. However, in practical scenarios, introducing surface magnetization on only one layer can generate a net Hall effect. Subsequently, our analysis is confined to the upper half of the system due to its symmetry.
}
The surface Chern number $\mathbb{C}_{z}$ is determined
by the sum of the local Chern markers ${C_{z}}(l)$
of the first several layers
\citep{varnava2018surfaces,gu2021spectral}.
Computational results indicate that ${C_{z}}(l)$
becomes nearly zero for $l>4$. Therefore, we consider  $\ensuremath{{\mathbb{C}_{z}}={\sum_{l=1}^{4}}{C_{z}}\left(l\right)}$
which is shown in Fig. \ref{fig:1}(b).
For a small magnetization $M_A$ (e.g. $M_{A}=0.5$),
a gapped Dirac cone occurs at the $\Gamma$ point [as
shown in Fig. \ref{fig:2}(c)], leading to
the local Chern markers $C_{z}(1)\approx 1/2$, $C_{z}(20>l>1)\approx 0$,
and the surface Chern number
$\mathbb{C}_{z} =1/2$ [See Figs. \ref{fig:2}(b) and \ref{fig:1}(b)].
This finding is well consistent with the results of previous theoretical
works from the low-energy models and has been observed in the experiments \citep{essin2009magnetoelectrica,mogi2022experimental,sekine2021axionelectrodynamics,nenno2020axionphysics,qi2008topological,fu2007topological,zhou2022transport}.
With the increase of the surface magnetization $M_A$, at $M_A$ slightly {less than} $4$,
the bands close and reopen at the X and Y points [Fig.
\ref{fig:2}(d)], leading to the local Chern marker $C_{z}(1)$ on the first layer
abruptly jumping to $-1$.
Meanwhile, $C_{z}(2)$ of the second layer sharply changes
to $-1/2$ [see the orange line in Fig. \ref{fig:2}(b)],
resulting in the surface Chern number $\mathbb{C}_{z} =-3/2$ [Fig. \ref{fig:1}(b)].
With $M_{A}=8$, after the bands close and reopen
at the M point [Fig. \ref{fig:2}(e)],
$C_{z}(1)$ abruptly becomes $0$,
while $C_{z}(2)$ remains approximately $-1/2$.
Consequently, $\mathbb{C}_{z} =-1/2$ [Fig. \ref{fig:1}(b)].
Fig. \ref{fig:2}(c-e) only presents
several instances under distinct magnetization $M_{A}$,
while a more comprehensive evolution of band structure
is documented in the Fig. \ref{fig:S2} of Appendix \ref{C}.

In the following, we will provide a physics interpretation of
the transition of surface Chern number $\mathbb{C}_{z}$
from $\frac{1}{2}$ to $-\frac{3}{2}$ to $-\frac{1}{2}$.
When the surface magnetization $M_A$ is relatively small,
without gap closing, adiabatic changes ensure that $\mathbb{C}_{z}$
remains at $\frac{1}{2}$ without alteration, which is consistent
with the low-energy models \cite{qi2008topological,fu2007topological}.
With the increasing of $M_{A}$, accompanied by
the gap closing and reopening at points X and Y,
the first layer enters the QAH phase.
Simultaneously, a massive Dirac fermion reemerges in the second layer.
This effect is analogous to the impact of Anderson disorders on the
surface of 3D TIs \citep{schubert2012fateof}, where the transition
of the surface QAH phase also leads to the revival of topological
surface states in the second layer.
However, due to the effective magnetization on the second layer
through second-order effects being opposite in direction
to the first layer (see Appendix \ref{B}),
this results in a $-\frac{1}{2}$ surface Chern number on the second layer
[as shown in the lower part of Fig. \ref{fig:1}(a)].
The system can be effectively modeled as a combination of a 3D TI
and a monolayer QAH system [see the inset in Fig. \ref{fig:1}(b) or
the lower part in Fig. \ref{fig:1}(a)],
leading to $\mathbb{C}_{z} = -1+(-\frac{1}{2}) = -\frac{3}{2}$.
As $M_{A}$ continues to increase, similarly, the gap closing and
reopening at point M corresponds to the transition of the first layer
from the QAH phase to the FMI phase, characterized by a local Chern
marker of zero. Consequently, due to
the coexistence of massive Dirac fermions and FMI monolayer
[the inset in Fig. \ref{fig:1}(b)],
the surface Chern number $\mathbb{C}_{z}$ becomes $0+(-\frac{1}{2})=-\frac{1}{2}$.
{ Additionally, it's worth mentioning that the essence of surface phase transition lies in the finite size of the Brillouin zone in real systems. Therefore, the occurrence of the phase transition is independent of the specific lattice model.}

\section{\label{sec:level3}Three layers effective model and phase diagram}

\begin{figure}
\begin{centering}
\includegraphics[scale=0.56]{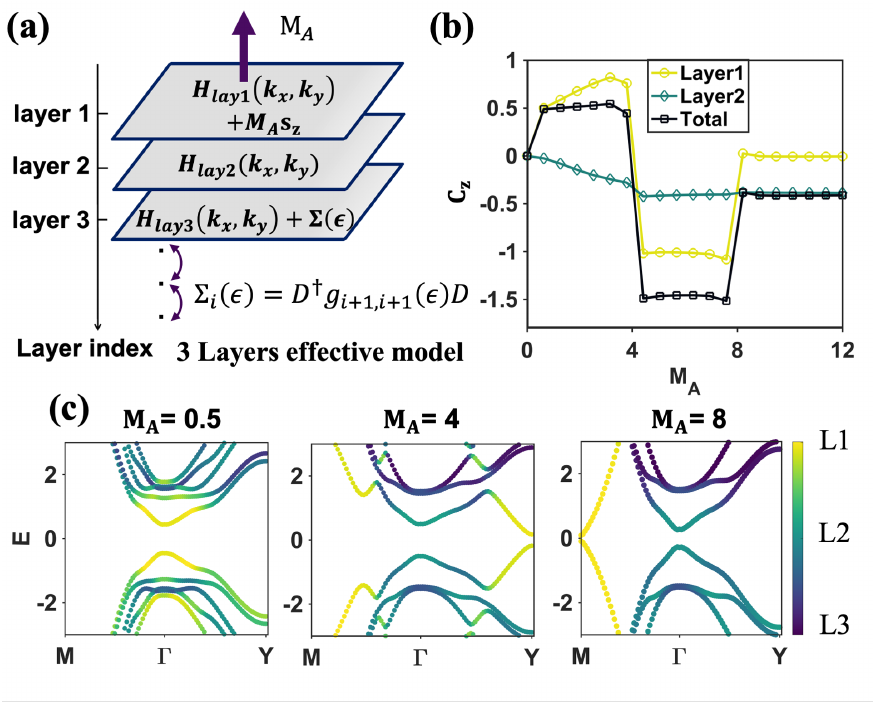}
\par\end{centering}
\caption{\label{fig:3}
(a) Schematic diagram of the three-layer effective
model. Ellipses {indicate} the process of self-energy iteration.
(b) $C_z(1)$, $C_z(2)$, and $\mathbb{C}_{z}$ as a function of $M_A$.
(c) Band structures under different $M_A$ with colors indicating
the layer of the states.
}
\end{figure}

To capture the surface physics of a 3D TI in the presence of surface
magnetization $M_A$, we employ a simplified three-layer effective model,
as illustrated in Fig. \ref{fig:3}(a).
The corresponding Hamiltonian is expressed as follows:
\[
\ensuremath{H=\left({\begin{array}{ccc}
{{H_{lay}}+{M_{A}}{s_{z}}} & D & 0\\
{D^{\dagger}} & {H_{lay}} & D\\
0 & {D^{\dagger}} & {{H_{lay}}+\Sigma\left(\varepsilon\right)}
\end{array}}\right)}
\]
Here the surface magnetization ${M_{A}}{s_{z}}$ is introduced
in the first layer, and the bulk's effective influence is
incorporated into the third layer through a self-energy term $\ensuremath{\Sigma\left(\varepsilon\right)}$.
The self-energy term $\ensuremath{\Sigma\left(\varepsilon\right)}$
can be obtained through iterative calculation of the Green's function
\citep{dattascattering}.
Specifically, for the i-th layer, the self-energy is defined as $\ensuremath{{\Sigma_{i}}\left(\varepsilon\right) ={D^{\dagger}}{g_{i+1,i+1}}\left(\varepsilon\right)D}$,
where ${g_{i+1,i+1}}\left(\varepsilon\right)= 1/\left({\varepsilon-{H_{lay}}-{\Sigma_{i+1}}\left(\varepsilon\right)}\right)$
represents the surface Green's function for the $i+1$th layer,
and $\varepsilon$ denotes energy.
Since the gap closing occurs at $\ensuremath{\varepsilon}=0$,
we consider $\ensuremath{\Sigma\left({\varepsilon=0}\right)}$ due
to its effective description of physics around $\ensuremath{\varepsilon}=0$.

We calculated the local Chern markers $C_z(l)$ under
varying surface magnetization $M_{A}$ [Fig. \ref{fig:3}(b)].
The computation results reveal a transition
of the surface Chern number $\mathbb{C}_{z}$
from $\frac{1}{2}$ to $-\frac{3}{2}$ to $-\frac{1}{2}$
[black line in Fig. \ref{fig:3}(b)], coinciding with the
numerical results shown in Fig. \ref{fig:1}(b).
It is noteworthy that, before the first gap closing
(at X, Y points), as $M_{A}$ increases,
the local Chern maker of the first layer gradually increases from
$\frac{1}{2}$ to 1, while that of the second layer decreases from 0.
Although there is a redistribution of $C_z(l)$,
the total Chern number of the three-layer system,
i.e. surface Chern number $\mathbb{C}_{z}$, remains at $1/2$.
This redistribution of $C_z(l)$ can be detected through
measurements of the ME coefficient, in Sec. \ref{sec:level4}, we will provide a detailed discussion of this.
Fig. \ref{fig:3}(c) displays the band structures and
layer-resolved projections under different $M_{A}$.
At $M_{A}=0.5$, the color of the Dirac cone at $\Gamma$ point
is yellow, indicating that the wave function is
predominantly distributed in the first layer.
The presence of a gapped Dirac cone on the first layer
corresponds to $C_z(1)\approx \frac{1}{2}$ at $M_{A}=0.5$ [Fig. \ref{fig:3}(b)].
However, with the increase of $M_{A}$,
the structure of the Dirac cone gradually disappears,
which corresponds with the redistribution of $C_z(l)$
between the first layer and the second layer.
After the gap closing and reopening at the X/Y point,
a Dirac cone structure reemerges in the second layer,
marked by green in the band structure [Fig. \ref{fig:3}(c) with $M_{A}=4$].
The reappearance of the Dirac cone in the second layer results
in $C_z(2)$ approaching $-\frac{1}{2}$
[see Fig. \ref{fig:3}(b)], indicating that the topological transition
in the first layer leads to the reformation of topological surface
states in the second layer and $\mathbb{C}_{z} \approx -\frac{3}{2}$.
With the furthermore increase of $M_A$,
as the band {closes and reopens} at the M point, the local Chern marker
of the first layer becomes zero, signifying the transition of the
first layer to the FMI phase and $\mathbb{C}_{z}$ jumping to $-\frac{1}{2}$.
The detailed evolutions of energy bands with changing magnetization
can be found in the Fig. \ref{fig:S2} of Appendix \ref{C}.

\section{\label{D} Phase diagram of 3D TI with surface magnetization}

\begin{figure}
    \begin{centering}
    \includegraphics[scale=0.8]{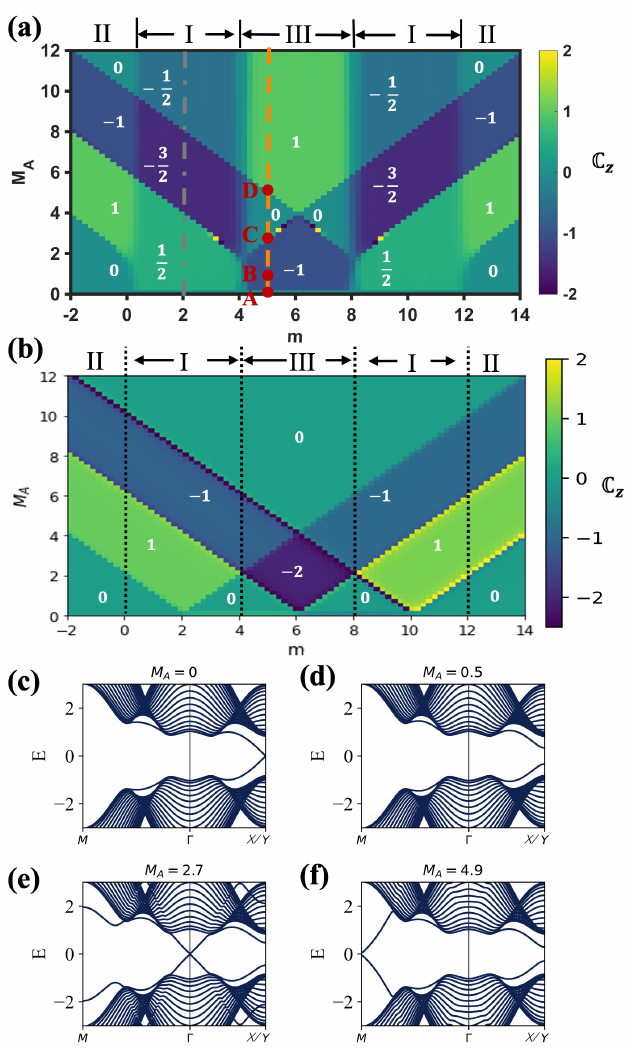}
    \par\end{centering}
    \caption{\label{fig:phase} (a) Surface Chern number $\mathbb{C}_{z}$ versus the surface magnetization
    $M_A$ and the mass term $m$. Fig. \ref{fig:1}(b) corresponds
    to the gray dash-dotted line in (a).
    (b) $\mathbb{C}_{z}$ of the single-layer model
    changes in response to alterations in $M_A$ and $m$.
    (c-f) These figures respectively correspond to the band
    structures at points A-D, which are identified with red markers in
    (a).}
    \end{figure}

Next, we investigate how the surface Chern number
$\mathbb{C}_{z}$ is affected by the topology of the bulk
which is defined by the topological mass term $m$ \citep{bernevig2013topological}.
Fig. \ref{fig:phase}(a) shows $\mathbb{C}_{z}$ with respect
to both the surface magnetization $M_{A}$ and the mass term $m$.
The transition of $\mathbb{C}_{z}$ from $\frac{1}{2}$ to
$-\frac{3}{2}$ to $-\frac{1}{2}$ can only be observed
in the topologically nontrivial bulk region (interval
I), i.e., $0<m<4B$ and $8B<m<12B$ \citep{shen2012topological}.
{ In }the topologically trivial bulk regions (II, III),
the behavior of $\mathbb{C}_{z}$ with respect to changes in
$M_{A}$ is inconsistent.
In the region II where $\ensuremath{m<0}$ and $\ensuremath{m>12B}$,
as $M_{A}$ increases, $\mathbb{C}_{z}$ changes
from $0\to1\to-1\to0$.
In the region III ($\ensuremath{4B<m<8B}$),
as $M_{A}$ increases, the surface Chern number $\mathbb{C}_{z}$ exhibits
a transition of $-1\to0\to1$.
We attribute the distinct behavior of
$\mathbb{C}_{z}$ in regions II and III to different $Z_{4}$ topological classification \citep{fu2007topological,shen2012topological,fu2007topological2}.

To provide an explanation for the differing behavior of surface Chern
number $\mathbb{C}_{z}$ in the regions II and III under the influence of surface
magnetization $M_A$, we calculated the phase diagram as a function of
the magnetization $M_A$ and topological mass $m$
for the single-layer system (See Fig. \ref{fig:phase}(b)).
This diagram reflects surface phase transitions under
magnetization conditions without bulk effects.
In region II, the variation of $\mathbb{C}_{z}$
in the single-layer system matches that of the multi-layer system,
implying that $\mathbb{C}_{z}$ of the multi-layer system
remains unaffected by bulk influence.

Conversely, in the topologically trivial region III,
changes in surface magnetization $M_A$ {result in different behaviors}
in the surface Chern number $\mathbb{C}_{z}$
between the single-layer and multi-layer systems.
In this case, $\mathbb{C}_{z}$
undergoes a transition from $0\to-2\to-1\to0$ in the single-layer system
[see Fig. \ref{fig:phase}(b)],
but $\mathbb{C}_{z}$ changes from $-1\to 0\to1$ in the multi-layer system
[see Fig. \ref{fig:phase}(a)],
indicating the impact of the bulk within the region III.
Specifically, at $M_A=0$, the multi-layer model exhibits
two Dirac cones at the X/Y points due to the bulk presence
[see Fig. \ref{fig:phase}(c)].
However, a single-layer system lacks these Dirac cones.
When the surface magnetization $M_A$ is small,
the two Dirac cones open up gaps [Fig. \ref{fig:phase}(d)],
each contributing $-1/2$ to surface Chern number $\mathbb{C}_{z}$.
Consequently, $\mathbb{C}_{z}$ of the multi-layer system
under small $M_A$ is $-1$.
As the surface magnetization $M_A$ increases,
$\mathbb{C}_{z}$ of the multi-layer system
in the region III transitions from $-1$ to 0.
The physics picture here is similar to the surface Chern number $\mathbb{C}_{z}$
transition from $1/2$ to $-3/2$ in the region I,
i.e., the topological transition of the first layer
and the reformation of Dirac cone in the second layer.
In particular, accompanied by the non-adiabatic change of the first
layer (gap closing and reopening at $\Gamma$ point, see Fig. \ref{fig:phase}(e)),
the first layer enters the QAH phase with $\mathbb{C}_{z}=-1$,
and simultaneously, the Dirac cone at the X/Y point re-forms
in the second layer.
Furthermore, the effective magnetization direction on the second
layer is opposite to that of the first layer, leading to gap opening
and contributing to $+1$ surface Chern number $\mathbb{C}_{z}$.
Therefore, $\mathbb{C}_{z}$ of the multi-layer system results
from the QAH of the first layer and
the newly formed gapped Dirac cones in the second layer,
yielding a total surface Chern number of $-1+1=0$.
Similarly, with further increases
in $M_A$, the first layer undergoes another non-adiabatic
change (gap closing and reopening at $M$ point, see Fig. \ref{fig:phase}(f)),
transitioning to a zero Chern number in the FMI phase.
Consequently, the overall surface Chern number $\mathbb{C}_{z}$
of the multi-layer system is contributed by the second layer,
resulting in $+1$.

In summary, {both region II and region III} fall under the topological trivial phase ($v_{0}=0$) according to $Z_{2}$ topological classification.
However, their $Z_{4}$ topological numbers are distinct \citep{fu2007topological,fu2007topological2}.
For the region II, $Z_{4}=(0;0,0,0)$, and for the region III, $Z_{4}=(0;1,1,1)$,
which corresponds to the presence of an even number of Dirac cones
on the surface of the region III \citep{shen2012topological}. Consequently,
we attribute the contrasting behavior of surface Chern numbers under
surface magnetization variations between the region II and region
III to their distinct $Z_{4}$ topological classifications.

\section{\label{sec:level4}Topological ME effect and six terminal Hall transport}

\begin{figure}
\begin{centering}
\includegraphics[scale=0.50]{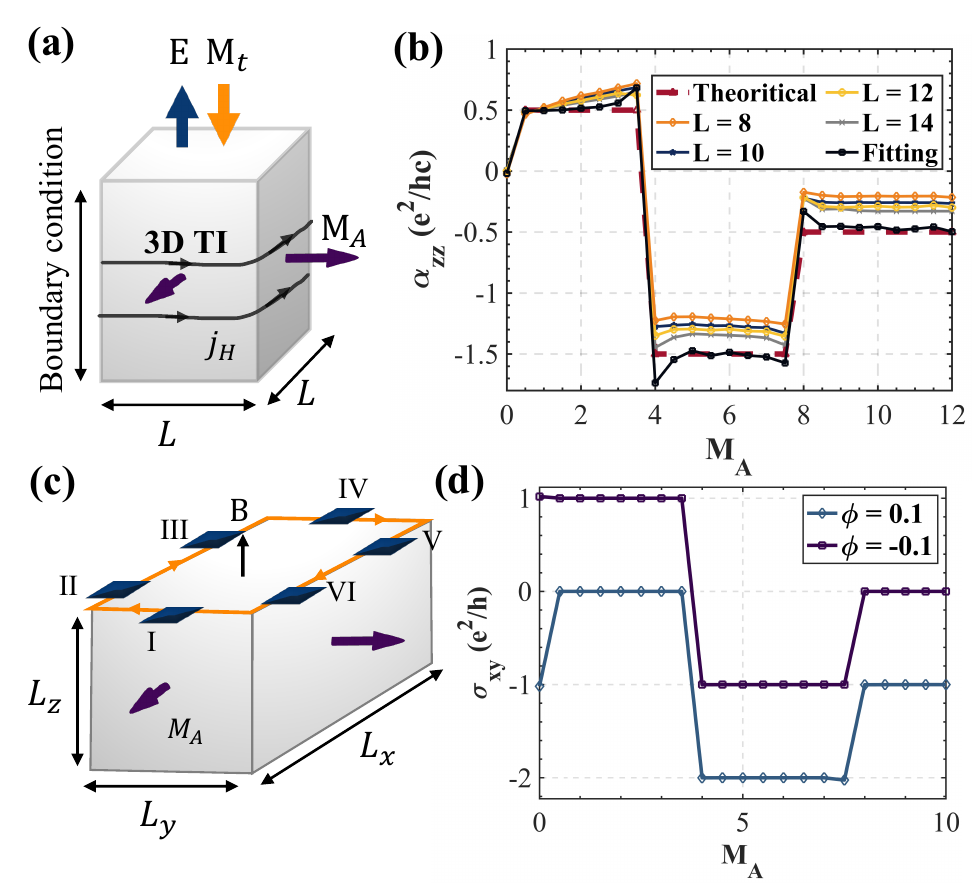}
\par\end{centering}
\caption{\label{fig:4} (a) Sketch of a 3D TI with surface magnetization $M_A$
normal to the side surfaces (purple arrows). With the application of an electric
field \ensuremath{\mathbf{E}}, the surface Hall effect induces a circulating
current denoted as $\ensuremath{{j_{H}}}$, consequently giving rise
to a bulk magnetization $\ensuremath{{{\bf {M}}_{{\bf {t}}}}\parallel{\bf {E}}}$.
(b) ME coefficient $\alpha_{zz}$ as a function of $\ensuremath{{M_{A}}}$
with different side length \ensuremath{L}.
(c) Illustration of a Hall bar device constructed using
a 3D TI, featuring metallic leads attached to the upper surface
(highlighted in blue).
The upper surface displays the chiral edge channels, as denoted by
the orange curve.
In this context, the magnetic field along the z-direction
and the magnetization on the lateral surface are represented by black
and purple arrows, respectively.
(d) Hall conductance on the upper surface as the function
of the side surface magnetization $M_{A}$ with
Fermi energy $\ensuremath{{E_{F}}=0.1}$.}
\end{figure}

The discontinuous platform transition of the surface Chern number $\mathbb{C}_{z}$,
driven by variations in surface magnetization $M_A$,
can be experimentally detected by measuring the ME response coefficient $\alpha_{zz}$ \cite{dziom2017observationa,vopson2017measurement}.
It's worth noting that, to detect the ME effect,
we apply outward-pointing magnetization $M_A$ on the lateral surface
of a 3D TI in the subsequent calculations, as depicted in Fig. \ref{fig:4}(a).
Upon applying an electric field $E_{z}$ in the z-direction,
the lateral surface will trigger a Hall current
$\ensuremath{{j_{H}}={\sigma_{H}}{E_{z}}}$, where
$\sigma_{H}$ represents the surface Hall { conductance}
[as shown in Fig. \ref{fig:4}(a)].
According to Ampere's law, the magnetization
along the z-direction is defined as $\ensuremath{M_{z}={j_{H}}/c}$,
where $c$ is the speed of light.
As the ME response coefficient satisfies the relation
$\ensuremath{{M_{z}}={\alpha_{zz}}{E_{z}}}$,
the ME coefficient and the lateral surface Hall conductance differ
by only a constant $c$ \citep{sekine2021axionelectrodynamics}.
Consequently, we expect that with an increase in
lateral surface magnetization $M_A$,
the system's ME response coefficient will exhibit a transition from
$\frac{1}{2}$ to $-\frac{3}{2}$ and then to $-\frac{1}{2}$,
as shown by the red dashed line in Fig. \ref{fig:4}(b).
To evaluate the ME effect, we employ linear
response theory to compute the orbital magnetization generated by
a vertically applied electric field $E_{z}$ (see Appendix \ref{F}).
By constructing a square in the \ensuremath{x}\ensuremath{y}-plane
with a side length of \ensuremath{L} sites
while maintaining translation symmetry along the \ensuremath{z}
direction, we obtain the ME response coefficient $\alpha_{zz}$ via
the Kubo formula \citep{wan2022topological,mahan2000manyparticle}.
The numerical results for the $\alpha_{zz}$ with different side length
\ensuremath{L} are illustrated in Fig. \ref{fig:4}(b).
As the $\alpha_{zz}$ only converges when \ensuremath{L} tends
to infinity \citep{rauch2018geometric}, increasing $L$ progressively
brings the numerical $\alpha_{zz}$ closer to theoretical values.
Additionally, we derive the ME response coefficient $\alpha_{zz}$
at the thermodynamic limit ($\ensuremath{L\to\infty}$) through fitting
(see Appendix \ref{F}),
and the fitting results shows agreement with the theoretical outcomes
[see Fig. \ref{fig:4}(b)].

As discussed in Sec. \ref{sec:level3} [refer to Fig. \ref{fig:3}(b)],
we have observed that for $M_{A}<4$, an increase in magnetization $M_A$
leads to a redistribution of local Chern markers across multiple layers of the surface.
However, since the overall surface Chern number remains constant,
this redistribution of local Chern markers may not be detectable
through conventional transport experiments.
However, we emphasize that the changes in the
distribution of local Chern markers can be experimentally observed
by measuring the ME response coefficient during variations in the
lateral surface magnetization [as shown in Fig. \ref{fig:4}(a)].
Specifically, in finite-sized systems where $M_{A}<4$,
as the lateral surface magnetization increases,
the ME response coefficient gradually rises [see Fig. \ref{fig:4}(b) ].
This can be attributed to the distinct cyclotron radii of the current
in the outermost and second outermost layers.
Consequently, this leads to different weights
in the contributions of the outermost and second outermost layers to
the z-direction magnetization.
In experiments, this phenomenon can
be observed by measuring the change in magnetic flux along the z-direction
using a Superconducting Quantum Interference Device (SQUID).

The variation of the surface Chern number $\mathbb{C}_{z}$
with the surface magnetization $M_A$ can also be observed
through distinctive Hall conductance plateau transitions
in a six-terminal system.
Let's consider a 3D TI slab under open boundary conditions
with the sizes $L_{z}=15$, $L_{y}=20$ and $L_{x}=100$
[Fig. \ref{fig:4}(c)].
We assume a magnetic field applied along the z direction,
corresponding to a magnetic flux of $\ensuremath{\phi=\pm0.1}$
per unit cell.
The surface magnetization $M_A$ is applied to the lateral surfaces
(but we still use $\mathbb{C}_{z}$ to describe the surface Chern number
of the lateral surface).
Employing a six-terminal Hall bar structure as depicted in Fig. \ref{fig:4}(c),
we calculate the Hall conductance using the Landauer-B$\ddot{u}$ttiker formula \citep{dattascattering,buttiker1988absence,fisher1981relation,landauer1970electrical,gong2020transport}(see
Appendix \ref{G}).
Fig. \ref{fig:4}(d) illustrates
how the upper surface Hall conductance changes with varying
lateral surface magnetization $M_{A}$ for $\ensuremath{\phi=\pm0.1}$.
When $\phi=0.1$, the magnetic field leads to $\mathbb{C}_{\phi}=\frac{1}{2}$
on the upper surface \citep{fu2007topological}.
As $M_{A}$ increases, a transition of surface Chern numbers $\mathbb{C}_{z}$
occurs on the lateral surface, transitioning from $\frac{1}{2}$ to $-\frac{3}{2}$
finally to $-\frac{1}{2}$.
The difference between $\mathbb{C}_{\phi}$ and $\mathbb{C}_{z}$
results in a domain wall between the lateral and upper surfaces, leading
to the chiral modes appearing at the boundary of the upper surface,
with the number of the chiral modes $\ensuremath{n=\mathbb{C}_{z}-\mathbb{C}_{\phi}}$.
The Hall conductance of the upper surface corresponds
to the number of chiral modes, i.e., $\ensuremath{\sigma_{xy}=n\frac{{e^{2}}}{h}}$.
Therefore, with the increasing $M_A$,
the Hall conductance undergoes a transition from $0$ to $-2\frac{e^2}{h}$ to
$-\frac{e^2}{h}$ [see the purple line in Fig. \ref{fig:4}(d)].
When $\phi=-0.1$, the Chern number on the upper surface is $-\frac{1}{2}$.
Similarly, with the increasing magnetization,
the Hall conductance undergoes a transition from $\frac{e^2}{h}$
to $-\frac{e^2}{h}$ to $0$,
as depicted by the purple line in Fig. \ref{fig:4}(d).

\section{\label{sec:level5}Discussion and conclusion}

{To investigate the possibility of realizing the topological phase transition mentioned in the paper using existing materials, calculations were performed based on the real parameters of the material $\text{Bi}_2 \text{Se}_3$ in Appendix \ref{real_mat}. However, the results indicate that the surface magnetic gap required for the phase transition is slightly larger than what can currently be achieved experimentally.}
In order to experimentally observe the surface phase transition of 3D TI,
it is required to minimize the material band width and maximize the Zeeman
splitting induced by the surface magnetization as much as possible.
Considering these requirements, 3D d-orbital TIs are ideal candidates
for realizing the aforementioned phase transition.
In such materials, the radial wave functions of the d orbitals
are predominantly distributed near the atomic nucleus, leading to reduced overlap integrals and narrower band widths.
Additionally, the Zeeman splitting effect of the d orbitals
is relatively strong.
A recent first-principles calculations
suggest that materials within the antifluorite $Cu_{2}S$ family can
achieve d-orbital strong 3D TI phase \citep{sheng2017dorbital}.
Furthermore, $Z_{2}$ strong TIs with flat band structures are
also an ideal choice for realizing such phase transitions
due to the small band width \citep{weeks2012flatbands}.
Meanwhile, quantum simulation techniques have undergone significant advancements
in recent years, and many condensed matter systems have achieved success
through this approach \citep{sugawa2018secondchern,ji2020quantum,xin2020quantum,cai2019observation,song2018observation,jotzu2014experimental,meng2016experimental,wu2016realization}.
Notably, Wilson lattice Hamiltonians, identical in form to the model used in this paper, have been realized in circuit systems \cite{yangRealizationWilsonFermions2023}. Therefore, quantum simulation methods also provide a powerful avenue
for investigating such surface phase transitions.

In summary, our study unveils a surface phase transition
with broken bulk-boundary correspondence
in 3D TIs induced by surface magnetization.
Employing both a realistic lattice model and a three-layer effective model,
we calculate the surface Chern number.
Remarkably, as surface magnetization increases, the surface
Chern number exhibits the transition from $\frac{1}{2}$
to $-\frac{3}{2}$ and finally to $-\frac{1}{2}$,
which cannot be explained by low-energy theories.
We attribute these transitions to the evolution
of the first-layer massive Dirac fermions into quantum QAH and FMI
phases, and the reemergence of a second-layer massive Dirac fermion
with opposite mass sign.
These topological transitions also lead to the sudden jumps in the
ME coefficient and the Hall conductance, which can
experimentally be observed.
Our investigation reveals a new phase with broken bulk-boundary
correspondence, enriching understandings of the properties of TIs.

\section{\label{sec:level6}Acknowledgment}
Y.-H. W. is grateful to Jiayu
Li, Ming Gong, Zhihao Huang and Ludan Zhang for fruitful discussions.
This work was financially supported by the National Natural
Science Foundation of China (Grant No. 12374034 and No. 11921005),
the Innovation Program for Quantum Science and Technology
(2021ZD0302403), and the Strategic priority Research
Program of Chinese Academy of Sciences (Grant No.
XDB28000000). We also acknowledge the Highperformance
Computing Platform of Peking University for
providing computational resources.

\appendix

\section{ \label{C}Evolution of Band Structure with Surface magnetization Variation}

\begin{figure*}
\begin{centering}
\includegraphics[scale=0.6]{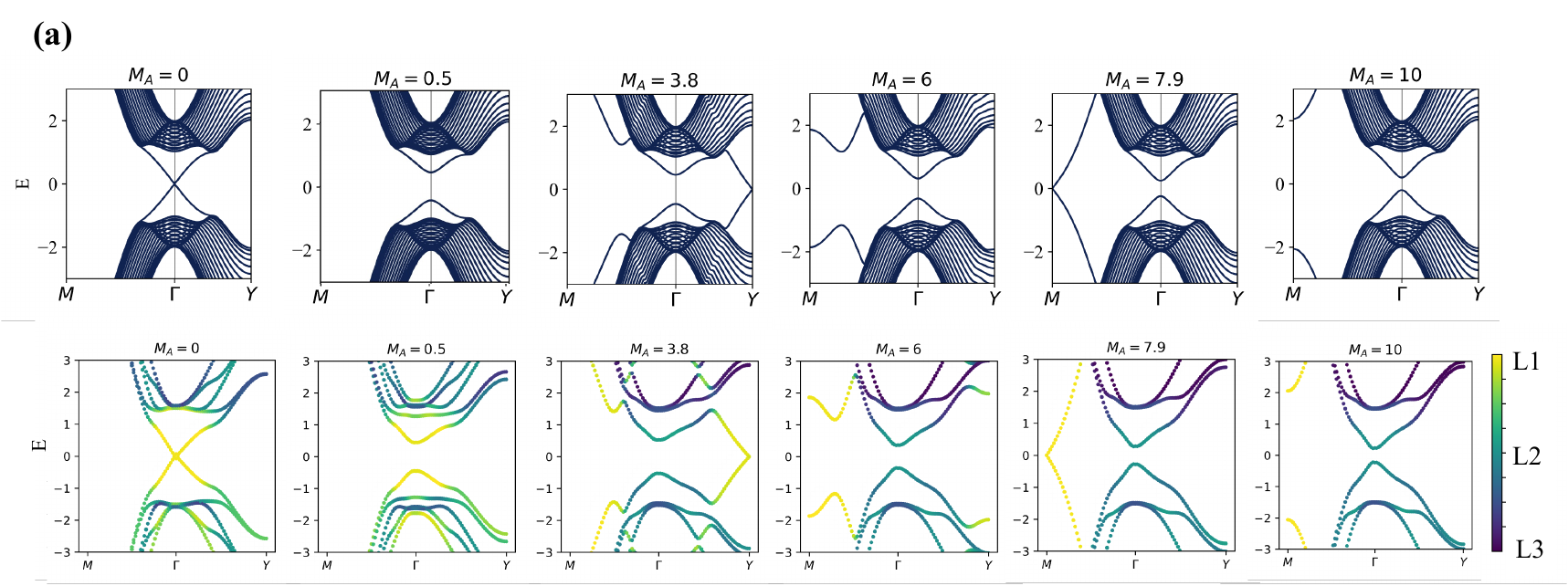}
\par\end{centering}
\caption{\label{fig:S2} (a) Band structure of systems under different
surface magnetizations $M_A$ (corresponding to the red points in Fig. \ref{fig:1}(b)).
Upper part: Band structure calculations for a 20-layer slab
model. Lower part: Band structure calculations for a 3-layer effective
model, where colors denote the state's corresponding layer.
The first, second, and third layers are respectively denoted by yellow, green,
and purple.}
\end{figure*}

In this section, we will present a more detailed variation of the
system's band structure with increasing surface magnetization $M_A$.
The model and computational methods used are consistent with those in
the main text. When the surface magnetization $M_{A}=0$, bulk-boundary
correspondence ensures the existence of gapless Dirac cones on the
surface of the 3D TI (see Fig. \ref{fig:S2} with $M_{A}=0$).
Upon introducing a small $M_A$, the breaking of TRS on the surface
leads to the gapped Dirac cone (Fig. \ref{fig:S2} with $M_{A}=0.5$)
and the surface Chern number $\mathbb{C}_{z}$ of 1/2 (Fig. \ref{fig:1}(b)).
Alongside the band closing and reopening at the XY points (Fig. \ref{fig:S2} with $M_{A}=3.8$),
the first layer undergoes a topological phase transition,
resulting in a jump of $\mathbb{C}_{z}$ from $1/2$ to $-3/2$.
With further increase in magnetization $M_A$,
the band at the M point experiences
a similar closing and reopening (See Fig. \ref{fig:S2} with $M_{A}=7.9$),
causing $\mathbb{C}_{z}$ to sharply
shift from $-3/2$ to $-1/2$.
From the projection of wave functions, we can observe that the $-1/2$
surface Chern number $\mathbb{C}_{z}$  is contributed by the massive
Dirac fermion in the second
layer (See Fig. \ref{fig:S2} with $M_{A}=10$).

\section{\label{B} Effective magnetization on second layer}

In our model, the magnetization $M_A$ is added
on the first layer (the outermost layer),
and the magnetization term is represented as $\text{\ensuremath{{M_{A}}{\sigma_{0}}{s_{z}}}}$.
However, an effective magnetization in the second layer
can be induced by the magnetization $M_A$ in the first layer.
The induced effective magnetization in the second layer can be obtained
by calculating the self-energy correction $\ensuremath{\Sigma_{mag}\left(\varepsilon\right)}$
of the first layer's magnetization
with respect to the second layer \citep{dattascattering},
\[
\ensuremath{\Sigma_{mag}\left(\varepsilon\right)=
{D^{\dagger}}\frac{1}{{\varepsilon-{M_{A}}{s_{z}}+i\eta}}D},
\]
where $\ensuremath{\eta}$ is an infinitesimally
small quantity approaching 0 and
$\text{\ensuremath{D=\frac{A}{{2i}}{\sigma_{x}}{s_{z}}+B{\sigma_{z}}{s_{0}}}}$
is the interlayer hopping with $A=1$ and $B=1$.
The symbol $\ensuremath{\varepsilon}$ represents
energy. Similar to the main text, here we adopt the approximation
$\ensuremath{\varepsilon=0}$, then the self-energy correction reduces into:
\[
\ensuremath{{\Sigma_{mag}}\left(0\right)=-\frac{5}{{4{M_{A}}}}{\sigma_{0}}{s_{z}}
+\frac{1}{{M_{A}}}{\sigma_{y}}{s_{0}}}
\]
The first term corresponds to the effective magnetization of the second
layer, with its sign opposite to that of the first layer.

\section{\label{F} Calculation and fitting of magneto-electric coupling coefficient}

\begin{figure}
\begin{centering}
\includegraphics[scale=0.35]{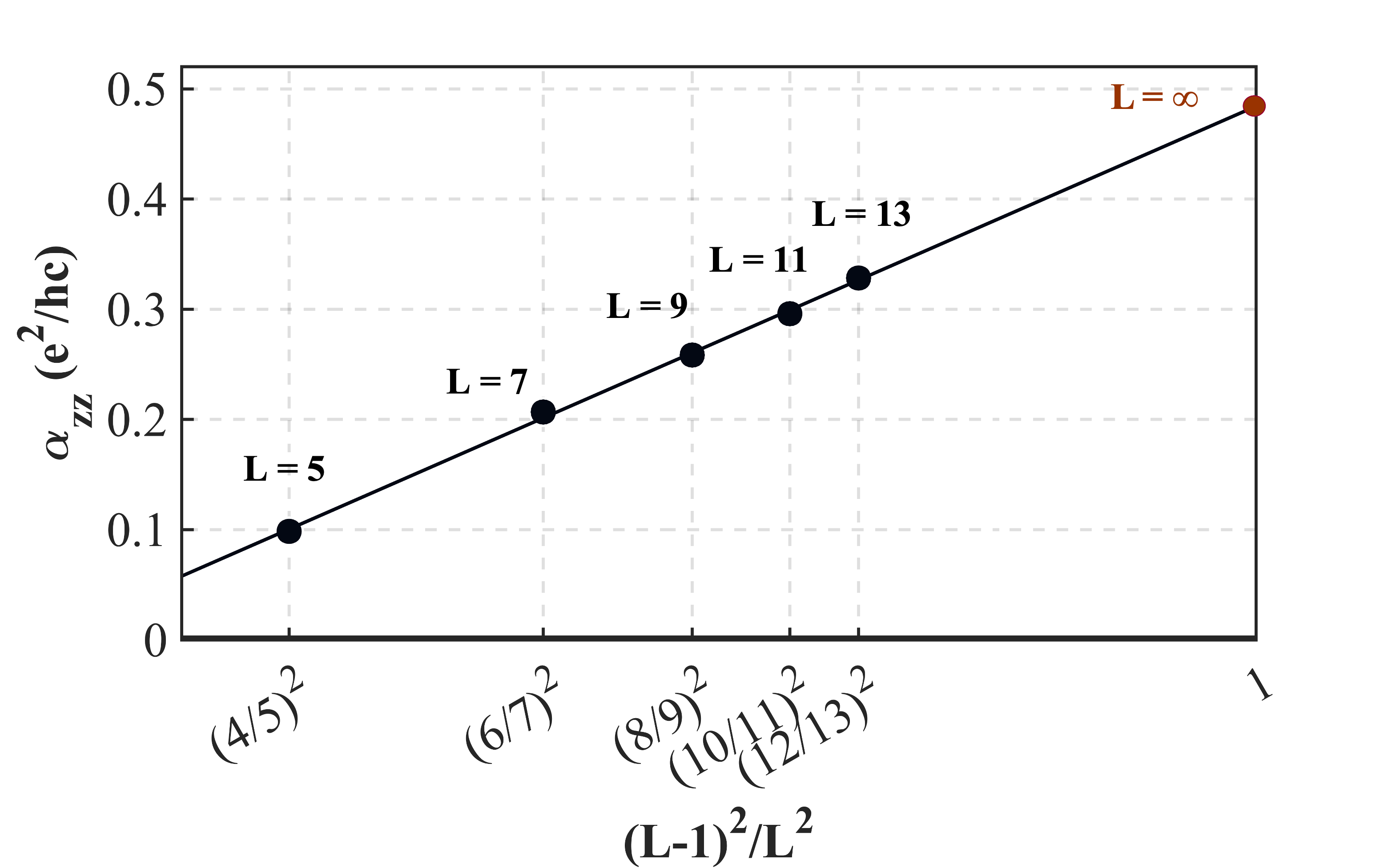}
\par\end{centering}
\caption{\label{fig:fitting} (a)
A representative diagram illustrates the ME coupling coefficient,
$\text{\ensuremath{\alpha_{zz}}}$, for systems with different side lengths $L$.
The ME coefficient in the thermodynamic limit ($\ensuremath{L\to\infty}$)
is marked by the red data point on the graph.
The surface magnetization $M_{A}=10.5$ in this plot.}
\end{figure}

ME response of a bounded sample can be derived
by linear response theory \citep{wan2022topological}:
\begin{equation}
    \alpha_{z z}= \sum_{\substack{k_z\\ i \in \text { occ. } \\ \mathrm{j} \in \text { unocc. }}} \frac{h\operatorname{Im}\left[\left\langle k_z, i\left|\hat{M}_z\right| k_z, j\right\rangle\left\langle k_z, j\left|\hat{J}_z\right| k_z, i\right\rangle\right]}{a N_{k_z} \mathscr{A}\left(\varepsilon_{i, k_z}-\varepsilon_{j, k_z}\right)^2}
    \end{equation}
The orbital magnetic moment, denoted as $\mathbf{M}$, is given by
$\ensuremath{\widehat{{\bf M}}=-(e/2c)\widehat{{\bf r}}\times\widehat{{\bf v}}}$
\citep{vanderbilt2018berryphases}. Here, $\widehat{{\bf r}}$ and
$\widehat{{\bf v}}$ represent the position and velocity operators,
and the current density $\mathbf{J}$ is expressed as $\mathbf{J}=e$
$\widehat{{\bf v}}$. $\mathscr{A}$ represents the cross-sectional
area in the xy-direction of the system, while $a$ stands for the
lattice constant in the z-direction. Here, $\ensuremath{\left|{{k_{z}},i}\right\rangle }$
denotes the i-th eigenstate with momentum $\ensuremath{{k_{z}}}$
and corresponding eigenvalue $\ensuremath{{\varepsilon_{i, k_z}}}$. In
the summation, $i$ and $j$ are limited to occupied and unoccupied
bands, respectively.

Magnetization on the lateral surface leads to a redistribution of
the local Chern markers between the outermost and second outermost
layers, causing z-directional electric fields to induce Hall currents
in both layers simultaneously.
Due to the diverse angular momentum of Hall ring currents in different layers,
variations occur in the electrically induced magnetic moment $M$,
subsequently causing changes in the $\alpha_{zz}$.
The coefficients $\ensuremath{\alpha_{zz}^{out/in}}$
are defined as the ME coupling coefficients of the system
considering only the magnetic moments generated by the Hall currents
in the outermost (second outermost) layer.
Since the overall magnetic moment of the system results from
the contributions of both the outermost and second outermost layer currents,
there exists a correlation between
$\alpha_{zz}$ and $\ensuremath{\alpha_{zz}^{out}}$, $\ensuremath{\alpha_{zz}^{in}}$.
    \begin{equation}
    L^{2}\alpha_{zz}=\ensuremath{L^{2}\alpha_{zz}^{out}+\left(L-1\right)^{2}\alpha_{zz}^{in}}
    \end{equation}
$\alpha_{zz}$ converges in the thermodynamic limit ($L\rightarrow\infty$) \citep{rauch2018geometric}
    \begin{align*}
    {\alpha_{zz}} & ={\lim_{{\rm {L}}\to\infty}}\left[\alpha_{zz}^{out}+{\left({\frac{{L-1}}{L}}\right)^{2}}\alpha_{zz}^{in}
    \right]\\
     & =\alpha_{zz}^{out}+\alpha_{zz}^{in}
    \end{align*}

To obtain $\alpha_{zz}$ in the thermodynamic limit, we calculate
$\alpha_{zz}$ for different side lengths $L$, and take $\alpha_{zz}^{out}+\alpha_{zz}^{in}$
as the fitting outcome for $\alpha_{zz}$ (see Fig. \ref{fig:fitting}).

    \section{ \label{G} Landauer-B$\ddot{u}$ttiker formula}

    In a Hall-bar configuration like that shown in Fig. \ref{fig:4}(c), we can employ
    the Landauer-B$\ddot{u}$ttiker formula to compute the Hall conductance on
    the system's surface. The Landauer-B$\ddot{u}$ttiker formula is expressed as:\citep{add1,add2}
    \begin{equation}
    \ensuremath{\left[{\begin{array}{lccccc}
    {T_{1}} & {T_{12}} & {T_{13}} & {T_{14}} & {T_{15}} & {T_{16}}\\
    {T_{21}} & {T_{2}} & {T_{23}} & {T_{24}} & {T_{25}} & {T_{26}}\\
    {T_{31}} & {T_{32}} & {T_{3}} & {T_{34}} & {T_{35}} & {T_{36}}\\
    {T_{41}} & {T_{42}} & {T_{43}} & {T_{4}} & {T_{45}} & {T_{46}}\\
    {T_{51}} & {T_{52}} & {T_{53}} & {T_{54}} & {T_{5}} & {T_{56}}\\
    {T_{61}} & {T_{62}} & {T_{63}} & {T_{64}} & {T_{65}} & {T_{6}}
    \end{array}}\right]\times\left[{\begin{array}{c}
    {V_{1}}\\
    {V_{2}}\\
    {V_{3}}\\
    {V_{4}}\\
    {V_{5}}\\
    {V_{6}}
    \end{array}}\right]=\left[{\begin{array}{c}
    {I_{1}}\\
    {I_{2}}\\
    {I_{3}}\\
    {I_{4}}\\
    {I_{5}}\\
    {I_{6}}
    \end{array}}\right]}\label{eq:3}
    \end{equation}
where $T_{p}= -\sum_{q (q\not= p)}T_{pq}$.
    $\ensuremath{{V_{p}}}$ and $I_{p}$ represent the voltage and current
    at lead $p$, respectively.
    The transmission coefficient from lead $q$ to lead $p$,
    denoted as $\ensuremath{{T_{pq}}}$, can be obtained
    using the formula $\ensuremath{{T_{pq}}=Tr\left({{\Gamma_{p}}{G^{R}}{\Gamma_{q}}{G^{A}}}\right)}$,
    where $\ensuremath{{G^{R/A}}={\left({{\varepsilon_{F}}-{H_{D}}-\sum_{p=1}^{6}\Sigma_{p}^{R/A}}\right)^{-1}}}$ are
    the retarded and advance Green's functions.\citep{add1,add2}
    $\ensuremath{{\Gamma_{p}}=i\left({\Sigma_{p}^{R}-\Sigma_{p}^{A}}\right)}$
    is the line width function, with $\ensuremath{\Sigma_{p}^{R/A}}$
    being the self energy of lead $p$.\citep{add1,add2}
    Given the current vector as $\ensuremath{\vec{I}{\rm {=[I,0,0,-I,0,0]}}}$,
    the voltage at each port can be determined by solving Eq. (\ref{eq:3}).
    The longitudinal resistivity $\ensuremath{{\rho_{xx}}=\left({{V_{2}}-{V_{3}}}\right)/I}$
    and Hall resistivity $\ensuremath{{\rho_{xy}}=\left({{V_{2}}-{V_{6}}}\right)/I}$.
    Furthermore, the Hall conductance can be computed using $\ensuremath{\sigma_{xy}=\rho_{xy}/\left({\rho_{xx}^{2}+\rho_{xy}^{2}}\right)}$.
    In the calculations, we set the central region's thickness as $\ensuremath{{L_{z}}=15}$,
    width as $\ensuremath{{L_{y}}=20}$, and length as $\ensuremath{{L_{x}}=100}$.
    The height and width of the leads are set as 3 and 10, respectively.

    {  \section{ \label{real_mat}Phase transition points under real material parameters}

In the paper, we utilize dimensionless parameters to enhance the clarity of the physical picture. To underscore practical relevance, this appendix focuses on utilizing parameters specific to realistic materials.

We compare our approach with the low-energy model of $\text{Bi}_2 \text{Se}_3$ \cite{add3}, where the Hamiltonian near the Gamma point is given by
\[
\begin{aligned}
    H^{3 \mathrm{D}} = & \epsilon_0(\mathbf{k}) s_0 \sigma_0 + M(\mathbf{k}) s_0 \sigma_z + A_1 k_z s_z \sigma_x \\
    & + A_2\left(k_x s_x + k_y s_y\right) \sigma_x.
\end{aligned}
\]
Here, ${k_ \bot } = k_x^2 + k_y^2$, $\epsilon_0(\mathbf{k}) = C + D_1 k_z^2 + D_2 k_{\perp}^2$, and $M(k) = M_0 - B_1 k_z^2 - B_2 k_{\perp}^2$. In this study, we simplify by setting ${\varepsilon _0}\left( k \right) = 0$, ${M_0} = 0.28\ eV$, ${B_1} = {B_2} = 0.1\,eV\,n{m^2}$, and ${A_1} = {A_2} = 0.1\,eV\,nm$. The parameters ${M_0}$ and ${B_{1}}$ here are consistent with the real material $\text{Bi}_2 \text{Se}_3$ \cite{add3}.

Based on this low-energy model within a cubic lattice with a lattice constant of $a=1 \ nm$ allows us to establish the relation between the parameters of real materials and those in Eq. (\ref{eq:1}).
 For realistic parameter of $\text{Bi}_2 \text{Se}_3$:${B_1}/{a^2} = 0.1 \ eV$, it corresponds to the dimensionless parameter $B = 1$ in Eq. (\ref{eq:1}). Similarly, the topological mass ${M_0} = 0.28 \ eV$ corresponds to $m = 2.8$, and ${A_1}/a = 0.1 \ eV$ corresponds to $A = 1$. Under this parameter mapping, the phase transition induced by the surface magnetization of $\text{Bi}_2 \text{Se}_3$ can be observed in the phase diagram  Fig. \ref{fig:phase}(a), 
 where the unit 1 corresponds to $0.1 \ eV$ for $\text{Bi}_2 \text{Se}_3$. With the increase of surface magnetization, the surface magnetic gaps corresponding to two phase transitions are approximately ${M_{A1}} \approx 0.32 \ eV$ and ${M_{A2}} \approx 0.72 \ eV$ respectively. However, the intrinsic magnetic topological insulator $\text{MnBi}_2\text{Te}_4$ currently exhibits the largest observed surface magnetic gap, measuring more than 60 meV \cite{estyunin2020signatures,otrokov2019prediction}. Therefore, achieving the phase transition discussed in this paper still presents some challenges with the existing materials.
    }

\bibliography{ref}

\end{document}